\documentclass[12pt]{article}
\usepackage{jheppub}
\usepackage{amssymb,amsfonts}

\usepackage{wrapfig}
\usepackage{epsfig}
\usepackage[T1]{fontenc}
\usepackage[utf8]{inputenc}
\usepackage{lmodern}
\usepackage{float}
\usepackage{placeins}

\usepackage[dvipsnames]{xcolor}
\usepackage{tikz}

\def\be{\begin{eqnarray}}
\def\ee{\end{eqnarray}}
\newcommand{\beq}{\begin{equation}}
\newcommand{\eeq}{\end{equation}}
\newcommand{\nn}{\nonumber}
\newcommand\para{\paragraph{}}

\newcommand{\eqn}[1]{(\ref{#1})}

\def\Dslash{\,\,{\raise.15ex\hbox{/}\mkern-12mu D}}
\def\Dbarslash{\,\,{\raise.15ex\hbox{/}\mkern-12mu {\bar D}}}
\def\delslash{\,\,{\raise.15ex\hbox{/}\mkern-9mu \partial}}
\def\delbarslash{\,\,{\raise.15ex\hbox{/}\mkern-9mu {\bar\partial}}}
\def\pslash{\,\,{\raise.15ex\hbox{/}\mkern-9mu p}}
\def\calDslash{\,\,{\raise.15ex\hbox{/}\mkern-12mu {\cal D}}}

\newcommand{\sign}{{\rm sign}}

\newcommand\bx{{\bf x}}

\newcommand{\numberset}{\mathbb}

\newcommand{\R}{\numberset{R}}

\newcommand{\Z}{\numberset{Z}}

\newcommand{\mc}[1]{\mathcal{#1}}

\newcommand{\cL}{\mc{L}}

\newcommand{\identity}{{1}}

\newcommand{\tr}{{\rm tr}}

\usetikzlibrary{decorations.markings, calc}

\tikzset{
    set arrow inside/.code={\pgfqkeys{/tikz/arrow inside}{#1}},
    set arrow inside={end/.initial=>, opt/.initial=},
    /pgf/decoration/Mark/.style={
        mark/.expanded=at position #1 with
        {
            \noexpand\arrow[\pgfkeysvalueof{/tikz/arrow inside/opt}]{\pgfkeysvalueof{/tikz/arrow inside/end}}
        }
    },
    arrow inside/.style 2 args={
        set arrow inside={#1},
        postaction={
            decorate,decoration={
                markings,Mark/.list={#2}
            }
        }
    },
}

\def\lae{\mathrel{\mathop{\smash{\lower .5 ex \hbox{$\stackrel<\sim$}}}}}
\def\lae{\mathrel{\mathop{\smash{\lower .5 ex \hbox{$\stackrel>\sim$}}}}}

\title{Evidence for a Non-Supersymmetric 5d CFT from Deformations of  5d  $SU(2)$ SYM}

\author{Pietro Benetti Genolini$^1$,}
\author{Masazumi Honda$^1$,}
\author{Hee-Cheol Kim$^2$,}
\author{David Tong$^1$,}
\author{and Cumrun Vafa$^3$}

\affiliation{$^1$Department of Applied Mathematics and Theoretical Physics, \\
${\ }$University of Cambridge, Cambridge, CB3 OWA, UK}
\affiliation{$^2$Department of Physics, POSTECH, Pohang 790-784, Korea}
\affiliation{$^3$Department of Physics, Harvard University, Cambridge, MA 02138, USA}

\abstract{We study supersymmetry breaking deformations of the ${\cal N}=1$  5d fixed point known as $E_1$, the UV completion of $SU(2)$ super-Yang--Mills.  The phases of the non-supersymmetric theory can be characterized 
by Chern--Simons terms involving  background $U(1)$ gauge fields, allowing us to identify a phase transition at  strong coupling.   We propose that this may signify the emergence of a non-trivial, non-supersymmetric CFT in $d=4+1$ dimensions.
}

\begin{document}
\pagestyle{plain} \setcounter{page}{1}
\newcounter{bean}
\baselineskip16pt \setcounter{section}{0}

\maketitle

\section{Introduction}

As the dimension of spacetime increases,  Gaussian fixed points have fewer relevant operators. This makes it increasingly difficult to start with a free theory and drive it to strong coupling in the infra-red. By the time we hit $d=4+1$ dimensions, we are out of options and we must take a more creative route if we are to discover interacting strongly coupled behaviour.

\para
One possibility, first mooted in \cite{peskin}, is to study Yang--Mills theory in the $d=4+\epsilon$ expansion. It is straightforward to see that the theory exhibits  UV fixed point for small $\epsilon$, but it is unclear if it remains trustworthy at $\epsilon=1$. (A lucid discussion of the results and pitfalls of this approach can be found in \cite{tim}.) More recently, a $6-\epsilon$ expansion has been employed to give  evidence for a $O(N)$ fixed point in five dimensions for sufficiently high $N$ \cite{fei}. This putative fixed point was subsequently explored using bootstrap methods \cite{boot,bookseller,booty,boot2}.

\para
Nonetheless, it remains true that the best understood fixed points in five spacetime dimensions have supersymmetry. These were first found using string theory arguments \cite{nati} and have been explored in great detail in the intervening years \cite{Morrison:1996xf,Douglas:1996xp,Intriligator:1997pq,ofer1,ofer2,DeWolfe:1999hj}.

\para
The existence of interacting supersymmetric fixed points suggests a very natural way to explore the landscape of 5d field theories:  we start from a supersymmetric theory in the UV and  deform by a relevant operator. Of course, if we wish to break supersymmetry -- and we do -- then we necessarily relinquish some control, and since our starting point is strongly coupled,  the suspicion is that it will be difficult to say anything about where we land up. Nonetheless, in recent years there has been some success at breaking supersymmetry in lower dimensions to derive dualities for strongly coupled, non-supersymmetric field theories, albeit dualities that were known previously  \cite{shamit1,shamit2,mecarl}. 
In particular, the authors of \cite{shamit1,shamit2} used information about the topological phases of gapped theories to argue that certain flows from a supersymmetric fixed point should land on  non-supersymmetric fixed points. Related topological arguments have also been used to explore the phase structure of 4d gauge theories by adding  soft supersymmetry breaking terms 
to  both ${\cal N}=1$ and $\mathcal{N}=2$ super-Yang--Mills  \cite{Konishi:1996iz,Aharony:2013hda,Cordova:2018acb, Wan:2018djl}.

\para
In this short note, we apply similar arguments to explore the phase structure of RG flows that emanate from the  five dimensional $E_1$ critical point, better known as the UV completion of $SU(2)$ ${\cal N}=1$ supersymmetric Yang--Mills \cite{nati}. We deform the theory by relevant operators that, at weak coupling, gap out both the scalar and the fermion, leaving behind only the $SU(2)$ gauge field. Nonetheless, we argue that (given certain assumptions described more fully below), at strong coupling, certain non-perturbative states remain gapless. 
We propose that these may point to the existence of a non-supersymmetric, interacting fixed point in 4+1 dimensions.

\section{The \texorpdfstring{$E_1$}{E1} Critical Point}

The $E_1$ fixed point was first identified by Seiberg  \cite{nati}. It can be thought of as the minimal UV completion of $SU(2)$ super-Yang--Mills, with no discrete theta angle\footnote{
In 5d theory with $SU(2)$ gauge group,
there are two choices of $\theta$ angle ($\theta =0$ or $\pi$) coming from the fact $\pi_4 (SU(2))=\mathbb{Z}_2$ \cite{Douglas:1996xp}.
}. The fixed point has symmetry
\be F = SU(2)_I \times SU(2)_R\nn\ee
Here $SU(2)_R$ is the R-symmetry shared by all theories with eight supercharges while $SU(2)_I$ is the global symmetry that gives the theory its enticing name. (This is the first in a sequence of theories with  $E_n$ global symmetry, and $E_1=SU(2)$.)

\para
The conserved current $J^a_\mu$, with $a=1,2,3$ the $SU(2)_I$ index, resides in a short conformal multiplet together with a number of other conformal primary operators. These can be constructed by acting with the supercharge $Q$ on the superconformal primary $\mu^{ia}$, yielding \cite{shiraz,yuji,cdi}

\be \mu^{ai} \ \stackrel{Q}{\longrightarrow}\ \psi^{am}_{\alpha} \ \stackrel{Q}{\longrightarrow}\  M^a, J^a_\mu \nn\ee
Here $i=1,2,3$ is an $SU(2)_R$ index. The fermionic operators $\psi$ also carry an $m=1,2$ $SU(2)_R$ index, now in the fundamental, as well as the $\alpha$  spinor index. Both the current and the scalar operator $M$ are $SU(2)_R$ singlets. The operators $\mu$, $\psi$ and $(M,J)$ have dimension $\Delta =3, 3.5$ and $4$ respectively. In what follows we will make use of both the relevant scalar operators $\mu$ and $M$ to deform the theory.

\para
The deformation by the  the scalar operator $M^a$ is well studied. We add
\be \delta{\cal L} = h^a M^a\label{susydef}\ee
This can be thought of as weakly gauging the $SU(2)_I$ flavour symmetry and giving an expectation value $h$ to the real scalar in the vector multiplet. Importantly, this deformation preserves supersymmetry, but breaks $SU(2)_I \rightarrow U(1)_I$. The $E_1$ fixed point then flows to supersymmetric Yang--Mills with gauge group $SU(2)$ and vanishing discrete theta angle. The low-energy physics is given by 
\be
\cL_{\rm YM} = \frac{1}{g^2}\tr\left( -\frac{1}{2}F_{\mu\nu}F^{\mu\nu} - \mc{D}_{\mu}\phi \mc{D}^{\mu}\phi  - i\bar{\lambda} \gamma^\mu \mc{D}_\mu \lambda + D^i D^i  + i\bar{\lambda} [\phi ,\lambda ]   \right) 
\label{sym}\ee
Here $\lambda$ is a symplectic Majorana spinor; we describe properties of this spinor in Appendix \ref{spinorsec}.  The scale of the IR gauge coupling is set by the relevant perturbation in the UV:  $|h|=1/2g^2$

\para
The surviving $U(1)_I\subset SU(2)_I$ symmetry is identified as the topological current in the low-energy theory,
\be
J^{\rm top}= \frac{1}{8\pi^2}*\tr \, F\wedge F \label{topJ}\ee
The fact that this topological symmetry is enhanced to $SU(2)_I$ at the fixed point was first noted in \cite{nati}, and has since been verified through analysis of  instanton zero modes \cite{yuji}, the superconformal index \cite{kimkimlee}, and the Nekrasov partition function \cite{mitev}. Indeed, the existence of such symmetry enhancement in the ultra-violet is a recurring theme in five dimensional gauge theories  \cite{Iqbal:2012xm,Bashkirov:2012re,Bergman:2013koa,Bergman:2013ala,Bergman:2013aca,Bao:2013pwa,Hayashi:2013qwa,Taki:2013vka,Aganagic:2014oia,Taki:2014pba,Hwang:2014uwa,Zafrir:2014ywa,Hayashi:2014wfa,Bergman:2014kza,Hayashi:2014hfa,Yonekura:2015ksa,ami,Apruzzi:2019opn}.

\para
The enhanced $SU(2)_I$ symmetry at the fixed point means that we flow to $SU(2)$ super-Yang--Mills regardless of the direction of the parameter  $h^a$ in \eqn{susydef}. In particular, if we fix a direction -- say $h^a = h \delta^{a3}$ -- then for both $h>0$ and $h<0$  we flow to $SU(2)$ super-Yang--Mills and, ultimately, to the free theory.

\begin{figure}[tb]
\begin{center}
\begin{tikzpicture}[scale=0.9]
    
	\draw [semithick] (-7,1) rectangle (-3,-1);	
	\draw [semithick] (-8,2) -- (-7,1);
	\draw [semithick] (-8,-2) -- (-7,-1);
	\draw [semithick] (-2,2) -- (-3,1);
	\draw [semithick] (-2,-2) -- (-3,-1);
	
	\draw [semithick] (3,2) rectangle (5,-2);
	\draw [semithick] (2,3) -- (3,2);
	\draw [semithick] (2,-3) -- (3,-2);
	\draw [semithick] (6,3) -- (5,2);
	\draw [semithick] (6,-3) -- (5,-2);
	
	\draw [dashed, thick, Blue] (-7,0) -- (-3,0) node [right=0.5cm] {\small instanton};
	\draw [dotted, thick, YellowOrange] (-5,1) -- (-5,-1) node [below=0.5cm] {\small W-boson};
	
	\draw [dotted, thick, Blue] (5,0) -- (3,0) node [left=0.5cm] {\small W-boson};
	\draw [dashed, thick, YellowOrange] (4,2) -- (4,-2) node [below=0.5cm] {\small instanton};
    
\end{tikzpicture}
\end{center}
\caption{Brane configurations corresponding to the pure $SU(2)$ gauge theory on the Coulomb branch. Horizontal lines represent D5-branes and vertical lines  NS5-branes. A fundamental string stretched between D5-branes corresponds to a W-boson in the field theory on the left-diagram, while it appears as an instanton of the dual $\widehat{SU(2)}$ gauge theory on the right. The supersymmetric CFT arises when the rectangle shrinks to a point.}
\label{fig:Branes}
\end{figure}
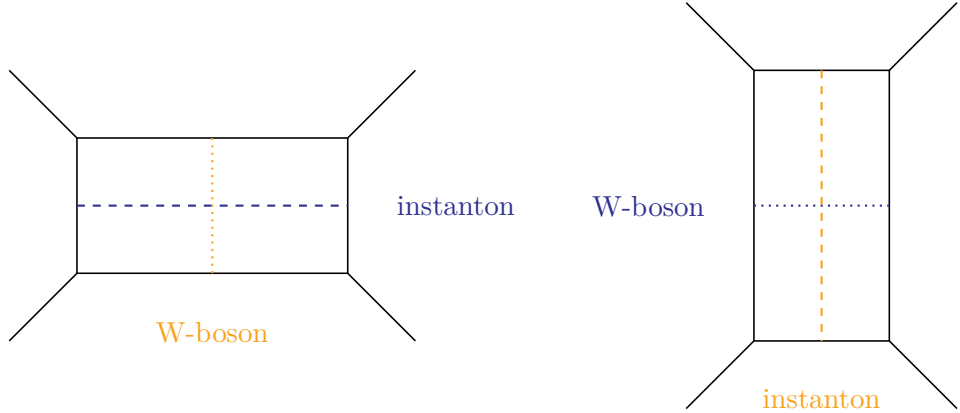

\para
If we move onto the Coulomb branch, then  the transition between the theories at $h>0$ and $h<0$ proceeds smoothly. This is  seen very clearly in the brane diagrams of  \cite{ofer1,ofer2}, as shown in the Figure \ref{fig:Branes}. Viewed from the low-energy field theory, this is a transition from a theory with $1/g^2>0$  into the regime that seemingly has $1/g^2<0$. The result can be viewed as a kind of UV duality, where the theory with $1/g^2<0$ is again described by super-Yang--Mills, but with a dual gauge group that we denote as $\widehat{SU(2)}$. The gauge couplings and scalar expectation values are related by
\be \frac{1}{\hat{g}^2} = -\frac{1}{g^2}\ \ \ {\rm and}\ \ \ \hat{\phi} =\phi + \frac{1}{2g^2}\label{duality}\ee
The W-bosons in one regime morph smoothly into the instantons in the other. 

\para
The supersymmetric conformal theory corresponds to the point where both $\phi,1/g^2\rightarrow 0$.  In this limit, the gluons and gluinos for both $SU(2)$ and the dual $\widehat{SU(2)}$ gauge group are massless.  The masses of the vector multiplets in $SU(2)$ are proportional to the vertical side of the rectangle in Figure \ref{fig:Branes} and the masses of the vector multiplets in $\widehat{SU(2)}$  are proportional to the horizontal side of the rectangle.  Moreover, when both gauge groups are massless in addition we get tensionless strings (which, on the Coulomb branch, arise as solitonic  monopole strings).  The tension of this string is proportional to the area of the rectangle in Figure \ref{fig:Branes}.  Note that when the $SU(2)$ gauge groups become massless, there are massless gluinos carrying $SU(2)_R$ charge.
Similarly, when $\widehat{SU(2)}$ is massless the vector multiplet states carry $SU(2)_I$ charge.  So, at the conformal point, we have massless modes carrying both $SU(2)_R$ and $SU(2)_I$ charges.  
\para
We will soon break supersymmetry and, in doing so, lift the Coulomb branch.  But the parameter $h$ which controls the coupling $ 1/g^2$ will continue to be important, as will the enhanced $SU(2)_I$ symmetry at the UV fixed point.  We will provide evidence that,  even after supersymmetry breaking,  massless modes carrying both $U(1)_R\subset SU(2)_R$ and $U(1)_I\subset SU(2)_I$ charges persist at infinite coupling.

\section{Breaking Supersymmetry}\label{susybreaksec}

The superconformal current multiplet contains a second relevant scalar operator, $\mu$, with dimension $\Delta[\mu]=3$. We can turn this on to flow away from the $E_1$ fixed-point, but only at the expense of breaking supersymmetry. This can be viewed as weakly gauging the $SU(2)_I$ flavour symmetry and giving an expectation value to the D-term in the vector multiplet.

\para
Our primary interest in this paper lies in the RG flows that result from turning on  both relevant operators at once\footnote{We discuss a different non-supersymmetric deformation in Appendix \ref{alternativesec}.},
\be \delta {\cal L} =  h^aM^a+ d^{ai}\mu^{ai} \ \ \ {\rm with}\ \ \ h^a = h\hat{v}^a\ \  {\rm and}\ \  d^{ai} = \tilde{m}^i \hat{v}^a\label{def}\ee
where $\hat{v}^a$ is a unit 3-vector. These deformations preserve a $U(1)_I\subset SU(2)_I$ as well as as the subgroup $U(1)_R\subset SU(2)_R$. We will be interested in the phase structure of the theory as we vary $h$ and $\tilde{m}$.
\para
When $|\tilde{m}|\ll h^2$,  we first flow to $SU(2)$ super-Yang--Mills \eqn{sym} and subsequently turn on a further mass deformation that breaks supersymmetry. This  mass deformation can be easily identified since it corresponds, up to a proportionality factor, to a turning on a D-term in a background $U(1)_I$ vector multiplet. The action \eqn{sym} is deformed by
\be \delta{\cal L} = m^i\,{\rm tr}\left(\frac{i}{4}\,\bar{\lambda}\sigma^i\lambda + \phi D^i\right)\label{break}\ee
where the IR deformation $m^i$ is proportional to the UV deformation $m^i\sim \tilde{m}^i$; we will see below, and in the appendix, that  this  proportionality factor includes a sign, so that $m^i = {\rm sign}(h)\,\tilde{m}^i$.
This gives a mass to $\phi$, lifting the Coulomb branch, as well as to the adjoint fermion $\lambda$. (The parameter $m^i$ has dimension 2; the physical mass of both the scalar and the fermion is $g^2m$.) The result is that the theory now flows to pure $SU(2)$ Yang--Mills in the infra-red.  We can, however, glean more information by studying the topological phase of the fermions. As we will see, this will ultimately allow us to also  say something about the strongly coupled phase $h^2 \ll |\tilde{m}|$.

\subsection{Topological Phases}

To make progress, we first make a choice  for the supersymmetry-breaking masses, say
\be \tilde{m}^i=(0,0,\tilde{m})\label{2def}\ee
This picks a specific choice of unbroken $U(1)_R\subset SU(2)_R$.   We then  introduce background gauge fields for our two global symmetries: $A_R$ for $U(1)_R\subset SU(2)_R$ and $A_I$ for $U(1)_I\subset SU(2)_I$. After integrating out the gapped fermions, we wish to determine the effective Chern--Simons term for these background fields
\be 
S_{CS} = \sum_{a=R,I} \frac{k_a}{24\pi^2} \int A_a\wedge dA_a\wedge dA_a  \label{csa}\ee
There can also be mixed Chern--Simons terms which we will discuss later in this subsection.

\begin{figure}[tb]
\begin{center}
\begin{tikzpicture}
    \tikzstyle{node_style}=[inner sep=2.5pt, circle]
    
	\draw [->] (-3,0) -- (3,0) node [right] {{\small $h$}};
	\draw [->] (0,-3) -- (0,3) node [above] {{\small $\tilde{m}$}};
    
    \node[node_style, fill=Blue, label=above left:{{\color{Blue}\small $E_1$}}] at (0,0) (E1) {};    
    
    \node[node_style, fill=MidnightBlue, label=right:{{\color{MidnightBlue}\small YM$_{(-2, -3/2)}$}}] at (2.5,2.5) (YMpTop) {};
    \node[node_style, fill=MidnightBlue, label=left:{{\color{MidnightBlue}\small YM$_{(+2, -3/2)}$}}] at (-2.5,-2.5) (YMpBottom) {};
    \node[node_style, fill=NavyBlue, label=left:{{\color{NavyBlue}\small YM$_{(+2, +3/2)}$}}] at (-2.5,2.5) (YMmTop) {};
    \node[node_style, fill=NavyBlue, label=right:{{\color{NavyBlue}\small YM$_{(-2, +3/2)}$}}] at (2.5,-2.5) (YMmBottom) {};

    \node[node_style, fill=Orange, label=below right:{{\color{Orange}\small SYM}}] at (2.5,0) (SYMR) {};
    \node[node_style, fill=Orange, label=below left:{{\color{Orange}\small SYM}}] at (-2.5,0) (SYML) {};
    
    \node[node_style, fill=BrickRed, label=above right:{{\color{BrickRed}\small ?}}] at (0,2.5) (UnknownTop) {};
    \node[node_style, fill=BrickRed, label=below left:{{\color{BrickRed}\small ?}}] at (0,-2.5) (UnknownBottom) {};
    
     \draw (E1) to [out=185,in=90] (YMpBottom) [arrow inside={end=stealth,opt={scale=2}}{0.4,0.75}];
     \draw (E1) to [out=-5,in=90] (YMmBottom) [arrow inside={end=stealth,opt={scale=2}}{0.4,0.75}];
     \draw (E1) to [out=5,in=-90] (YMpTop) [arrow inside={end=stealth,opt={scale=2}}{0.4,0.75}];
     \draw (E1) to [out=175,in=-90] (YMmTop) [arrow inside={end=stealth,opt={scale=2}}{0.4,0.75}];
     \draw (E1) to (SYMR) [arrow inside={end=stealth,opt={scale=2}}{0.4,0.75}];
     \draw (E1) to (SYML) [arrow inside={end=stealth,opt={scale=2}}{0.4,0.75}];
     \draw (E1) to (UnknownTop) [arrow inside={end=stealth,opt={scale=2}}{0.4,0.75}];
     \draw (E1) to (UnknownBottom) [arrow inside={end=stealth,opt={scale=2}}{0.4,0.75}];
\end{tikzpicture}
\end{center}
\caption{Phase diagram for 5d $SU(2)$ SYM; the subscripts denote the levels $(k_I,k_R)$ of the background Chern--Simons terms.  The  dark blue point at the origin is the strongly-coupled UV fixed point with enhanced global symmetry. Turning on relevant deformations triggers RG flows with different endpoints which, at weak coupling, coincide with pure Yang--Mills.  (Strictly speaking, the labels YM and SYM tell us about the physics close to the fixed point; the fixed point itself is free.)
The $\Z_2\times \Z_2$ symmetry of the diagram is due to the ``UV duality.''}
\label{fig}
\end{figure}
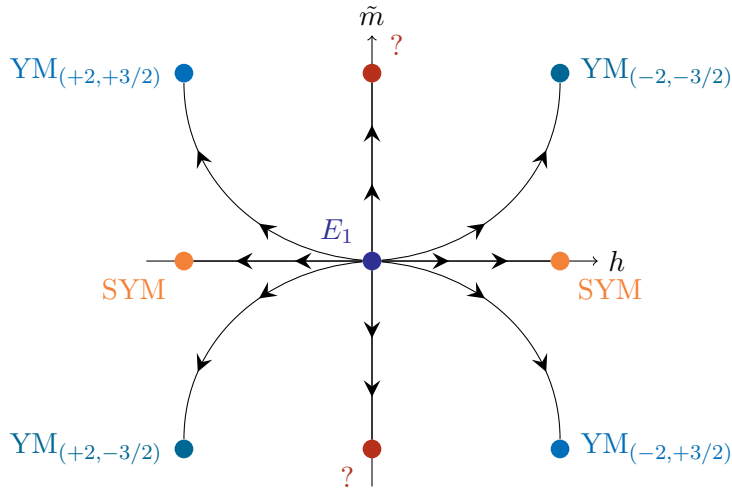

\para
Our goal is to determine the levels $k_R$ and $k_I$ in various parts of the phase diagram, labelled by $h$ and $\tilde{m}$.  This phase diagram is shown in Figure \ref{fig} and naturally splits into quadrants, depending on the sign of $h$ and $\tilde{m}$.  At a generic point in the phase diagram, the global symmetry of the theory is $U(1)_R\times U(1)_I$; this is enhanced to $SU(2)_R\times U(1)_I$ along the $h$-axis, except at the origin where it is further enhanced to $SU(2)_R\times SU(2)_I$.

\para
Crucially, if we determine the Chern--Simons levels in one quadrant --- say, $h>0$ and $\tilde{m}>0$ --- then we can determine them in all regions. This follows from the existence of a $\Z_2\times \Z_2$ symmetry acting on the moduli space of the theory, in which we act with $SU(2)_I$ and $SU(2)_R$ to continuously rotate the vector $\hat{v}^a$ and $\tilde{m}^i$ in \eqn{def} to their negative values. Acting with $SU(2)_I$ results in the map
\be
(h,{\tilde m})\rightarrow (-h,-{\tilde m}); \quad A_I\rightarrow -A_I ; \quad (k_I,k_R)\rightarrow (-k_I,k_R);
\label{tra1}
\ee
Acting with $SU(2)_R$ gives
\be
(h,{\tilde m})\rightarrow (h,-{\tilde m});  \quad A_R\rightarrow -A_R;  \quad (k_I,k_R) \rightarrow (k_I,-k_R);
\label{tra2}\ee
In particular,  combining these two operations we learn that, for a fixed ${\tilde m}$, as we cross the $h$ axis from $ h>0$ to $ h<0$ both levels flip sign:  $(k_I,k_R)\rightarrow (-k_I,-k_R)$.  Our task now is to evaluate  these Chern--Simons levels.

\para
The Chern--Simons term for $A_R$ can be determined by a simple perturbative calculation in the weakly coupled regime $h^2\gg |m|$. We work in the regime $h>0$.  The symplectic Majorana fermion $\lambda$ decomposes into fermions which carry charge $\pm 1$ under $U(1)_R$. Integrating out these fermions\footnote{This result is the same as for a massive Dirac fermion charged under $U(1)$. Details of the calculation for a Dirac fermion can be found, for example, in \cite{Bonetti:2013ela}. The computation for  a symplectic Majorana fermion is broadly similar, differing only in minor points. 
}
 induces the $A_R$ Chern--Simons term in \eqn{csa} with  
\be k_R = -\frac{3}{2}{\rm sign}(m)\label{kR}\ee
The familiar ${\rm sign}(m)$  term is consistent with the expectation (\ref{tra2}) based on symmetry. The factor of $3$ arises because $\lambda$ transforms in the adjoint of the $SU(2)$ gauge group.  The half-integer value for  $k_R$ reflects the fact that $SU(2)_R$ suffers a non-perturbative anomaly \cite{Intriligator:1997pq}; with $SU(2)_R$ broken to $U(1)_R$, this non-perturbative $\Z_2$ anomaly manifests itself as the familiar parity anomaly.

 \para
Next we turn to the background Chern--Simons terms for $U(1)_I$. There are no perturbative states carrying these quantum numbers so we learn nothing from simply integrating out the massive gluino. Nonetheless, there is a simple argument that fixes the level $k_I$. This follows from the requirement that the supersymmetric prepotential ${\cal F}$ is invariant under the UV duality \eqn{duality}, which sends 
\be h\rightarrow -h\ \ \ {\rm  and}\ \ \ \phi \rightarrow \phi+h\label{ddoesthis}\ee
This tells us that the prepotential (when $\tilde{m}=0$) takes the form\footnote{The prepotentials of rank 1 and rank 2 5d SCFTs with mass deformations are obtained in \cite{Hayashi:2019jvx} by using UV symmetries as we did here. The prepotential presented in (2.6) in \cite{Hayashi:2019jvx} for the $E_1$ theory agrees with our prepotential after the replacement $m_0\rightarrow4h$.}
\be 6{\cal F} = 12 h \phi^2 + 8\phi^3  - 2h^3\nn\ee
where the first term arises from the tree-level action, the second from a one-loop computation, and the final term is fixed by the duality. (The lack of an $h^2\phi$ can be argued as follows: any fermion charged under $U(1)_I$ must come with $\pm$ charges under $U(1)_{\rm gauge}$ and therefore contributes schematically as $|h + \phi| + |h-\phi|$. But, at weak coupling, $h\gg \phi$, this implies there is no $h^2\phi$ term. Moreover, there is no transition as the prepotential is extended to $h=0$.) The $h^3$ term in the prepotential contains the information  about the $U(1)_I$ Chern--Simons term, which we learn is 
\be k_I = -2\, {\rm sign}(h)\label{kI}\ee
The fact that the level depends on the sign of $h$ is consistent with (\ref{tra1}).

\subsection*{Evidence for a Non-Supersymmetric CFT}

As we pass from one quadrant to another in the $(h,{\tilde m})$ plane, the background Chern--Simons levels jump. This ensures that something interesting happens on each of the coordinate axes.  

\para
This ``something interesting'' could come in different flavours. Perhaps the least interesting something is that a symmetry is spontaneously broken. For example, there could be a phase at strong coupling in which the $U(1)_R$ and $U(1)_I$ symmetries are spontaneously broken. Alternatively, the $\Z_2$ symmetry which maps $h\rightarrow -h$ may be  spontaneously broken at $h=0$, resulting in a first order phase transition. We cannot rule out such scenarios.

\para
Nonetheless, under the assumption that the various global symmetries survive,  the jump in the Chern--Simons levels signifies the existence of new  massless modes, charged under the corresponding global symmetry. Such behaviour is seen if we fix $h$, and vary $\tilde{m}>0$ to  $\tilde{m}<0$, moving from the upper-right quadrant to the lower-right  in Figure \ref{fig}. Here the new light degrees of freedom are obvious: they are the massless scalar and gluino that emerge at the $\tilde{m}=0$ supersymmetric axis. The fermions are charged under $U(1)_R$ and neutral under $U(1)_I$, and this is in evidence in jump of the Chern--Simons levels.

\para
However, our analysis also shows that something interesting must happen on the $h=0$ axis as we fix $\tilde{m}>0$ and vary $h$ from positive to negative, transitioning from the upper-right quadrant to the upper-left in Figure \ref{fig}. If the global symmetries are not spontaneously broken then there must be massless modes. Since both $k_R$ and $k_I$ jump as we cross the $h=0$ axis, these modes must be charged under both symmetries. In particular, the fact that these modes are charged under $U(1)_I$ means that non-perturbative states become massless even after breaking supersymmetry. 
 
 \para
 This suggests that turning on the relevant, supersymmetry breaking operator $\mu^{ia}$, with $h=0$ in \eqn{def} results in a flow to fixed point with new massless degrees of freedom. If so, the important question becomes: what is the nature of this fixed point? Is it free? Or is it interacting?  We note  that in the supersymmetric theory, the existence of massless modes carrying $SU(2)_R$ and $SU(2)_I$ charges is the hallmark of an interacting conformal field theory. Relatedly, it is natural to conjecture that massless modes carrying both $U(1)_R$ and $U(1)_I$ charges signify a non-supersymmetric interacting fixed point. It would be very interesting to try to use bootstrap methods, along the lines of \cite{boot,bookseller,booty}, to look for evidence for the existence of such a non-supersymmetric CFT in 5d. 

\subsection*{Mixed Chern--Simons terms}

We can glean further information about the possible massless states that appear as we vary $m$ and $h$ by evaluating various mixed Chern--Simons terms. 
The simplest such terms arise for the $U(1)_R\times U(1)_I$ symmetries that we considered previously. In general, the Chern--Simons terms take the form,
\be S_{\rm mixed} = \sum_{a,b,c=R,I}\frac{k_{abc}}{24\pi^2} \int A_a\wedge dA_b\wedge dA_c\nn\ee
In this notation $k_{RRR}=k_R$ and $k_{III}=k_I$; these were computed in \eqn{kR} and \eqn{kI} respectively.  The mixed terms cannot be determined by a direct perturbative calculation because each involves the $U(1)_I$, under which only non-perturbative states are charged. Nonetheless, these too can be fixed by using the UV duality, as we now explain.

\para
First, the term with $k_{RII}$ vanishes when $\tilde{m}=0$. This is because this term is linear in $A_R$, but the full $SU(2)_R$ is unbroken when $\tilde{m}=0$. 
When we turn on small $\tilde{m}$, so that  $h^2\gg |m|$ and the theory is weakly coupled,  the masses of fermions charged under $U(1)_I$ are determined by the sign of $h$. Such fermions sit in weakly broken $SU(2)_R$ representations, and integrating them out cannot generate a mixed  $U(1)_R-U(1)_I^2$ Chern--Simons term. This ensures that $k_{RII}=0$ at weak coupling $h^2\gg |m|$.

\para
To compute the level $k_{RRI}$ we again employ the duality on the Chern--Simons terms in the supersymmetric theory. When $\tilde{m}=0$ and $\phi\neq0$, so we sit on the Coulomb branch,  the low-energy theory will have a mixed $U(1)_{\rm gauge}\times SU(2)^2_R$  Chern--Simons term 
\be
    S_{RRg}=\frac{q}{8\pi^2}\int A_g\wedge {\rm tr}(F_R\wedge F_R)\quad {\rm with}\quad q = 2\, {\rm sign}(\phi)
\ee
The supersymmetry relates this to an additional coupling of the form $2\phi (F_R)^2$ up to a numerical factor. 
Then the duality implies that the low-energy theory should involve another term $h (F_R)^2$ so that the linear couplings in $\phi$ and $h$ remain invariant under the duality map \eqn{ddoesthis}.  Supersymmetry then relates this to the mixed $U(1)_I\times SU(2)^2_R$ Chern--Simons level. After subsequently turning on $\tilde{m}$, so that $SU(2)_R$ is broken to $U(1)_R$, we have 
\be
    k_{RRI} = {\rm sign}(h)
\ee
Once again, this Chern--Simons level holds in the weak coupling regime $h^2 \gg |m|$, where there instanton states are all heavy. Note that, once again, this Chern--Simons level distinguishes the $h>0$ and $h<0$ phases.

\para
Finally, there are also mixed $U(1)$-gravitational Chern--Simons terms. These take the form
\be S_{\text{grav}} = \sum_{a=g,R,I}\frac{\kappa_a}{192\pi^2} \int A_a\wedge {\rm Tr} \left( R \wedge R \right) \nn\ee
where the sum is now over $U(1)_{\rm gauge}$, $U(1)_R$, and $U(1)_I$ symmetries, the former holding only on the Coulomb branch with $\phi\neq 0$.  The first two of these follow from standard perturbative calculations,
\be \kappa_g = 2\,{\rm sign}(\phi)\ \ \ {\rm and}\ \ \ \kappa_R =  -\frac{3}{2}{\rm sign}(m) \nn\ee
We can  then compute the mixed  $U(1)_I$-gravitational Chern--Simons term by relating it to $\kappa_g$, using the same kind of argument involving supersymmetry and the UV duality that we invoked to determine $k_{RRI}$. This time, we have
\be \kappa_I = {\rm sign}(h)\nn\ee

\subsubsection*{String Defects}

The different topological phases can also be seen in the behaviour of string defects. As we now explain, such string defects necessarily carry chiral fermions, where the chirality is determined by the sign of the Chern--Simons terms. 

\para
When such string defects are aligned along the $x^1$ direction, the background field strengths $F_a= dA_a$, with $a=R,I$, have a profile which obeys the modified Bianchi identity
\be dF_a = 2\pi q_a \prod_{i=2}^4 \delta(x^i)dx^i \label{bianchi}\ee
where $q_a$ labels the magnetic charge of the string. In the presence of such a string defect, a general, mixed Chern--Simons term transforms  under the gauge transformation $\delta A_a=d\Lambda_a$ as \cite{Ferrara:1996hh}
\be \delta S_{\rm cs} = \frac{k_{abc}}{8\pi^2} \int d(\Lambda_a)\wedge F_b \wedge F_c = -\frac{k_{abc}}{2 \pi} q_a \int_{\mathbb{R}^2}\Lambda_b F_c \label{inflow}\ee
This means that there exists anomaly inflow toward the 2d worldsheet of the string defect.

\para
The usual anomaly inflow argument means that this anomaly is cancelled by chiral modes on the defect. Typically, this happens if the chiral modes realise a $U(1)_R\times U(1)_I$ current algebra with level $|k_{ab}|$, where
\be k_{ab} = -k_{abc}q^c\nn \ee
This means that the 't Hooft anomalies can be computed from chiral modes living on the string defect.
We refer the reader to \cite{anomaly-inflow} and references therein for more details about the anomaly inflow in 5d gauge theory.  

\para
For  the  $U(1)_I$ string defect, the chirality of the zero modes on the string is dictated by the sign of the corresponding Chern--Simons term: $k_I = 2\,{\rm sign}(h)$. This means that there is a jump in the chirality of the zero modes as we cross from $h>0$ to $h<0$, again signalling the presence of a phase transition. 

\para
There, however, is a subtlety  for the $U(1)_R$ string defect with $q_R$ odd. This  arises because $U(1)_R$ symmetry suffers a parity anomaly, manifested by the half-integer Chern--Simons level \eqn{kR}. It is not possible to realise a chiral current algebra on the defect worldsheet carrying such a half-integer level. This means that the anomaly inflow argument cannot work for string defects with $q_R$ odd. Instead, the string defect provides a situation in which the $U(1)_R$ symmetry is not conserved. In Appendix \ref{spinorsec}, we exhibit a realisation of the string defect for the $U(1)_R$ symmetry and show that its worldsheet houses Majorana--Weyl fermions, which are neutral under $U(1)_R$.  We will show that these Majorana--Weyl fermions again flip chirality as we vary $h>0$ to $h<0$.

\para
Note that there is no such issue for $q_R$ even, where the chiral fermions are now expected to  furnish a representation of the current algebra. Alternatively, one could consider doubling the theory, and introducing a string defect for the diagonal $U(1)_R$ symmetry. The Majorana--Weyl fermions that we describe the in Appendix now come in pairs and again give a representation of the current algebra.

\appendix{}

\section{Appendix: Majorana Spinors and Zero Modes}\label{spinorsec}

The spinor representation of  $Spin(1,4)$ is pseudo-real. This does not allow us to impose a Majorana condition on a single fermion. However, if we take two Dirac fermions and ask that they transform in a doublet of an $SU(2)$ global symmetry, then this too is a pseudo-real representation.  This means that we can  impose a reality condition on a pair of Dirac fermions. The result is the symplectic Majorana spinor. It has the same number of degrees of freedom as a Dirac spinor, but with a manifest $SU(2)$ symmetry which, in the context of our supersymmetric theory, is identified with $SU(2)_R$.

\para
We will need to understand the properties of these symplectic Majorana fermions in some detail. We work in signature $(-++++)$ with gamma matrices
\be \gamma^0=\left(\begin{array}{cc} 0 & 1 \\ -1 & 0\end{array}\right)\,,\ \ \gamma^1=\left(\begin{array}{cc} 0 & 1 \\ 1 & 0 \end{array}\right)\,,\ \ \gamma^2=\left(\begin{array}{cc} \sigma_1 & 0 \\ 0 & -\sigma_1 \end{array}\right)\, ,\ \ \gamma^3=\left(\begin{array}{cc} \sigma_3 & 0 \\ 0 & -\sigma_3 \end{array}\right)\,,\ \ \gamma^4=\left(\begin{array}{cc} -\sigma_2 & 0 \\ 0 & \sigma_2\end{array}\right)\nn\ee
with $\sigma_a$ the usual Pauli matrices.

\para
We take parity to act as $x^1\mapsto -x^1$. Under parity and (anti-unitary)  time reversal, a Dirac fermion $\psi$ transforms as
\be {\cal P}:\psi \mapsto i\gamma^1 \psi\ \ {\rm and}\ \ {\cal T}:\psi \mapsto -i \gamma^0\gamma^4 \psi\label{pt}\ee
These obey ${\cal P}^2 = {\cal T}^2 = ({\cal P}{\cal T})^2 = (-1)^F$; these generate the quaternionic group $Q_8$.
Under charge conjugation, a Dirac spinor transforms as
\be {\cal C}: \psi \mapsto \psi^C = \gamma^4\psi^\star\label{charge}\ee
As anticipated above, it is not consistent to set $\psi = \psi^C$. Instead, we introduce to a pair of Dirac spinors, $\psi^m$, $m=1,2$, and impose the symplectic Majorana condition
\be \psi^m = \epsilon^{mn} (\psi^n)^C\label{SMajorana}\ee
(Strictly speaking, this should be called a {\it pseudo}-symplectic Majorana condition.)   For such a Majorana fermion there is no charge conjugation because $SU(2)_R$ is pseudoreal. 
(Indeed, the transformation \eqn{charge} now coincides with a rotation by $\pi/2$ in $SU(2)_R$.)

\para
Adding the supersymmetry breaking deformation breaks this global symmetry group $G$.  The naive, $SU(2)_R$-invariant mass term $\bar{\psi}\psi$ vanishes for a symplectic Majorana spinor. Instead, we have mass terms transforming as a triplet of $SU(2)_R$, 
\be {\cal L}_m = im^i (\bar{\psi}^m \sigma^i_{mn}\psi^n)\label{symplectic}\ee
These are the form of the  mass terms that arise in our supersymmetry breaking deformation \eqn{break}.  For any choice of $m^i$, the $SU(2)_R$ symmetry is broken to $U(1)_R$.  

\para
The mass deformation preserves a choice of time reversal\footnote{For the choice of mass $m^1$ or $m^3$, ${\cal T}'$  coincides with  
${\cal T}$. For $m^2\neq 0$, ${\cal T}$ is broken but we can twist with a broken element of $SU(2)_R$ to define a new ${\cal T}'$.}
${\cal T}'$ with $({\cal T}')^2 = (-1)^F$.  In contrast, the mass terms break parity ${\cal P}$. Since we now have a $U(1)_R$ symmetry, it is possible to define a new charge conjugation symmetry ${\cal C'}$, although one can check that this too is broken by the mass terms. However, the combination ${\cal C}'{\cal P}$ survives and obeys $({\cal C}'{\cal P})^2 =(-1)^F$ and ${\cal C}'{\cal P}{\cal T}' = (-1)^F{\cal T}'{\cal C}'{\cal P}$; together these generate the group $D_8$.

\para
There is a simple argument that  a massive spinor in 5d must break ${\cal C}'$ and ${\cal P}$. The little group in 5d is $SO(4) = SU(2)_l \times SU(2)_r$. Quantising a minimal spinor  in 5d (either Dirac or symplectic Majorana) gives rise to 4 states with vanishing momentum and these sit in the $(2,0)_+\oplus (0,2)_-$ representations of  $SO(4)\times U(1)_R$. This spectrum is invariant under neither parity  (which flips $SU(2)_l$ and $SU(2)_r$), nor under charge conjugation (which flips $+$ and $-$). But the spectrum is invariant under the combination ${\cal C}'{\cal P}$. These symmetries also tally with the induced Chern--Simons term \eqn{csa} which, in 5d, is odd under charge conjugation (which maps $A_R\rightarrow- A_R$) and parity, but even under time reversal (which maps $F_{0i}\rightarrow F_{0i}$ and $F_{ij}\rightarrow -F_{ij}$).

\subsection*{Domain Walls}

In Section \ref{susybreaksec}, we characterised the topological phase of the fermions by the level of the Chern--Simons term for a background $U(1)_R$ gauge field. It is also simple to see effect of this topological classification by considering a spatially dependent mass  $m(x^4)$,  which interpolates between $m<0$ at $x^4\rightarrow -\infty$ and $m>0$ at $x^4\rightarrow +\infty$.  
We take the mass to be aligned as in \eqn{2def}, 
\be m^i = (0,0, m(x^4))\nn\ee
In the presence of such an interface, the Dirac equation becomes
\be \delslash\psi^m  -  m(x^4)\, (\sigma_3)^{mn} \psi^n=0\ \ \ \Rightarrow\ \ \ \delslash\psi - m(x^4) \, \psi = 0\nn\ee
where, in the second equation, we have imposed the symplectic Majorana condition \eqn{symplectic} and written $\psi^1 \equiv \psi$. It is simple to check that the zero mode of this Dirac equation is a Weyl fermion in $d=3+1$ dimensions. This zero mode is protected by the time reversal symmetry ${\cal T}'$ (and, for a single Dirac fermion, by the $U(1)_R$ symmetry) and cannot be lifted.

\subsection*{String Defects}

We now describe another  manifestation of the phase transition as we vary  $h$ from positive to negative values. In particular, we show that the gapless modes on a string defect flip chirality as we cross the $h=0$ line. This follows from the general arguments involving string defects and anomaly inflow presented in Section \ref{susybreaksec}; here we flesh this out with more detailed calculations.

\para
We again consider turning on supersymmetry breaking mass deformations in the UV,  but this time we turn on a spatially dependent mass profile $m^i(x)$. Such mass profiles were employed long ago as a signal of topological phases (see, for example, \cite{ryureview}) and revisited more recently in the context of higher form symmetries \cite{cord}.

\para
Specifically, we allow the mass to wind in the spatial $\R^3$ parameterised by $(x^2,x^3,x^4)$, 
\be \tilde{m}^i(\bx) = \frac{m(r)}{r} y^i \ \ \ {\rm with}\  y^i = (x^2,x^3,x^4)\label{winding}\ee
with $m(r)$ a profile function that depends on $r^2 = (x^2)^2 + (x^3)^2 + (x^4)^2$ such that $m(r)\rightarrow 0$ as $r\rightarrow 0$ and $m(r)\rightarrow m$ as $r\rightarrow \infty$. This defines a defect in which the mass $m^i$ has  winding $+1$ in $\R^3$. The origin of the defect is $\R^{1,1}$ parameterised by $x^a$ with $a=0,1$. 
 In other words, this corresponds to a string defect. Note that the $SU(2)_R$ symmetry is twisted with the normal bundle from the $SU(2)_{\rm rot}$ rotation symmetry. We  can then further include a monopole profile \eqn{bianchi} for the background $U(1)_R$ gauge field $A_R$, so that we have  a 't Hooft--Polyakov monopole. However, for the zero mode counting of interest, we need only the winding \eqn{winding}.

\para

We again study this system in the weakly coupled regime $h>0$ with $h^2 \gg m$, where the physics is captured by mass-deformed super-Yang--Mills.  We decompose the 5d spinor as 
\be \psi(x,y) = \chi (x) \otimes \lambda (y)\nn\ee
Accordingly, we split Cliff$(1,4)\cong$ Cliff$(1,1)\, \otimes\, $Cliff$(3)$. Our choice of gamma matrices $\gamma^\mu$ decompose as 
\be
\gamma^a &=& \rho^a\otimes \identity_2 \ \ \ {\rm with}\ \ \rho^0 = i\sigma_2 \, , \quad \rho^1 = \sigma_1 \, , \quad \rho_\star = \rho^0\rho^1 = \sigma_3
\nn\\
 \gamma^{i+1} &=& \rho_\star\otimes \tau^i  \ \ \ {\rm with}\ \ \tau^1 = \sigma_1 \, , \quad \tau^2 = \sigma_3 \, , \quad \tau^3 = -\sigma_2
\nn\ee
The Dirac equation then becomes
\be
\frac{1}{g^2}\left( \rho^a \partial_a\chi \right) \otimes \lambda + \frac{1}{g^2}\left( \rho_\star\chi \right) \otimes \tau^i\partial_i\lambda - \chi^{C_2}\otimes \left[ \left( \tilde{m}_1 - i \tilde{m}_2\right) \sigma_2 \lambda^\star \right] - \tilde{m}_3 \, \chi \otimes \lambda = 0 \nn\ee
We seek solutions with $\chi$  a two-dimensional Majorana--Weyl zero mode, obeying 
\be
\rho^a \partial_a \chi = 0 \, , \quad \rho_\star\chi = \pm \chi \, , \quad  \chi = \chi^{C_2} \equiv \sigma_3\chi^\star 
\nn\ee
The resulting equation for the $3d$ spinor becomes
\be
\frac{1}{g^2} \tau^i \partial_i \lambda_\pm \mp \left[ (\tilde{m}_1 - i\tilde{m}_2) \sigma_2\lambda^\star_\pm + \tilde{m}_3 \lambda_\pm \right] = 0 \nn\ee
For the mass defect \eqn{winding} with winding $+1$, there is no normalisable solution for $\lambda_-$. There is, however,  a single normalisable solution for $\lambda_+$,
\be
\lambda_+ =  \exp \left( - g^2 \int_0^r dt\ m(t) \right) \begin{pmatrix}
1-i \\ 1+i 
\end{pmatrix}
\label{chiral}\ee
This corresponds to a right-moving Majorana--Weyl zero mode, $\rho_\star \chi=+\chi$ propagating along the defect. In contrast, if we take a mass defect with winding $-1$, we get a left-moving zero mode, obeying $\rho_\star\chi =-\chi$. These are  closely related to the zero modes discussed in \cite{jeff}. 

\para
In the context of super-Yang--Mills, the 5d fermion $\lambda$ transforms in the adjoint of the $SU(2)$ gauge group. 
Correspondingly, the Majorana--Weyl fermion zero mode on the defect also transforms in the adjoint of the bulk $SU(2)$ gauge group.

\para
As we noted in the main text, the fermi zero modes are not charged under $U(1)_R$, reflecting the $\Z_2$ parity anomaly in this symmetry. In contrast, we would expect to find the fermions on the charge $q_R=2$ string to carry $U(1)_R$ charge. Relatedly, if we were to instead double the theory and introduce a string for the combined $U(1)_R$ symmetry, this too would house 6 Majorana--Weyl fermions, or 3 Weyl fermions, which carry a $U(1)_R$ current algebra at level $3/2+3/2=3$.

\para
Once again, we ask: what happens to this system as we vary $h$ from positive to negative? This time we cannot rotate the vector $\hat{v}^a$ in \eqn{2def} since this would involve also rotating the spatial plane $\R^3$. Instead, we can make use of the outer automorphism  of $SU(2)$. From the brane-web picture, this operation is S-duality of IIB string theory; it was also applied in the field theoretic context to study duality walls in   \cite{dualitywall}.   The outer automorphism has the same effect, mapping
\be (h,\tilde{m}^i) \rightarrow (-h,-\tilde{m}^i)\nn\ee
But the physics remains the same. If we start in the weakly coupled regime $h\gg |\tilde{m}|$ with a defect exhibiting, say, a right-moving fermion then, after the duality transform, we must remain with a right-handed fermion. Yet the duality flips $\tilde{m}^i\rightarrow -\tilde{m}^i$ and hence flips the winding number in the UV. Since the winding is correlated with the chirality of the zero mode, the infra-red supersymmetry-breaking deformation $m^i$ must be related to the UV deformation by  $m^i \sim \sign(h)\,\tilde{m}^i$.

\para
This means that if we fix the winding of $\tilde{m}^i$ in the UV, and vary the relevant deformation $h$ from positive to negative, then we will transition from mass-deformed SYM with  a right-moving zero mode, to mass-deformed SYM with a left-moving zero mode. Clearly this cannot happen in a continuous fashion: there must be a phase transition.

\para
At weak coupling, the only way in which chiral zero modes can be lifted is if they become non-normalisable. This requires that the appropriate fermions become gapless in the bulk. A simple example of this arises if we generalise our discussion slightly. We could imagine a deformation which gives a mass to the fermion but, in contrast to \eqn{break}, leaves the scalar gapless, preserving (at least at the classical level) the Coulomb branch. The Dirac equation for the adjoint fermion $\lambda$ is then
\be \gamma^\mu {\cal D}_\mu \lambda - [\phi, \lambda] - \frac{1}{4}g^2 m^i \sigma^i \lambda=0\nn\ee
The fermion now receives a mass from both $m^i$ and the vacuum expectation value of the adjoint scalar $\phi$. 
It is simple to show that when these two are tuned so that 
$g^2 |m|=4|\phi|$, 
gapless modes emerge; they carry charges $(\pm 1,\mp 1)$ under the unbroken $U(1)_{\rm gauge}\times U(1)_R$. We can now examine what becomes of chiral zero modes when either $m^i$ or $\phi$ winds in some way. For example, in the presence of a mass defect, the single chiral zero \eqn{chiral} persists provided that 
$|\phi|<g^2 |m|/4$. 
However, it becomes non-normalisable\footnote{One can ask similar questions in the presence of a monopole string. When $m=0$, the monopole string houses two chiral zero modes, reflecting the fact that the worldsheet preserves ${\cal N}=(0,4)$ supersymmetry. These zero modes persist when a small mass is turned on. However, when we cross the threshold 
$g^2 |m|>4|\phi|$, 
the zero modes once again become non-normalisable. (See, for example, the appendix of  \cite{deboer} for an index theory analysis of this phenomenon.).} when the gapless mode appears in the bulk, and the mass defect has no zero modes for 
$|\phi|>g^2 |m|/4$.

\para
It is, of course, less clear what kind of transition we might have at strong coupling. Once again, we cannot rule out a first order transition in which some higher energy state, with zero modes propagating in the opposite direction, becomes the ground state.  Suffice to say that there is no remnant of such  of an excited, chiral state at weak coupling. 

\para
However, a more tantalising possibility is that the flip of chirality in the defect can be traced to the existence of tensionless strings, which themselves carry chiral fermionic modes. Such tensionless strings are themselves a hallmark of 5d supersymmetric fixed points. Clearly it would be interesting to understand if they also play a role in the putative non-supersymmetric fixed point.

\section{An Alternative Non-Supersymmetric Deformation}\label{alternativesec}

We note that there is a slightly different way to deform the $E_1$ fixed point which breaks supersymmetry and also results in something interesting in the infra-red.  This occurs if we set $h=0$ in \eqn{susydef} and turn on the deformation
\be \delta L = d^{ai}\mu^{ai} \ \ \ {\rm with}\ \ \ d^{ai} = \beta \delta^{ai}\label{su2flow}\ee
This breaks $SU(2)_I\times SU(2)_R\rightarrow SU(2)_{\rm diag}$. For this deformation, we can use anomaly matching arguments to get some understanding of where we're likely to end up. First note that $SU(2)_R$ has a discrete 't Hooft anomaly. This anomaly is associated to $\Pi_5(SU(2)) = \Z_2$, as  explained in \cite{Intriligator:1997pq}.  (It is closely related to the $\Pi_4(SU(2))=\Z_2$ Witten anomaly in four dimensions \cite{wittensu2}.) The anomaly can be seen in the super-Yang--Mills theory \eqn{sym} where there is a single symplectic Majorana fermion transforming in $SU(2)_R$, and therefore also exists at the fixed point. Meanwhile, there is no such anomaly for $SU(2)_I$. 
This means that $SU(2)_{\rm diag}$ inherits the anomaly and it must be present at the end of the flow \eqn{su2flow}.  We learn that either $SU(2)_{\rm diag}$ is spontaneously broken or there are gapless fermionic degrees of freedom that remain. The end point of this flow is therefore a candidate for a fixed point with $SU(2)_{\rm diag}$ symmetry. (It may, of course, simply be a free fermion.) Again, it would certainly be interesting to understand this further. 

\para
In addition, there are obvious generalisations of  these ideas to other higher rank UV fixed points that exhibit an enhanced $SU(2)_I$ symmetry. These include the UV completions of $SU(N)$ with Chern--Simons level $|\kappa| =N$ and $Sp(N)$ with an anti-symmetric hypermultiplet.
Indeed, given the classification proposal for 5d ${\cal N}=1$ supersymmetric theories  \cite{Bhardwaj:2019fzv} (some of which have no gauge theory origin), we have a large list of possible starting points for deformations leading to non-supersymmetric CFT's in 5d.
%

\section*{Acknowledgements}

We would like to thank Houri-Christina Tarazi and Ashvin Vishwanath for valuable discussions and participation at various stages of this project.
We have also benefited from discussions with Michele Del Zotto, Thomas Dumitrescu, Gary Gibbons, Kimyeong Lee, Shiraz Minwalla and Juven Wang. HK would like to thank Harvard University for hospitality during part of this work.
The research of PBG, MH and DT is supported  by the STFC consolidated grant ST/P000681/1. 
The research of HK is supported by
the POSCO Science Fellowship of POSCO TJ Park Foundation and the National Research Foundation of Korea (NRF) Grant 2018R1D1A1B07042934. DT is a Wolfson Royal Society Research Merit
Award holder and is supported by a Simons Investigator Award. 
The research of CV is
supported in part by the NSF grant PHY-1719924 and by a grant from the Simons
Foundation (602883, CV).

\end{document}